\documentstyle[twocolumn,aps,graphicx]{revtex}

\def\be{\begin{equation}}
\def\ee{\end{equation}}
\def\ba{\begin{eqnarray}}
\def\ea{\end{eqnarray}}

\begin{document}

\hyphenation{Ka-pi-tul-nik}

\twocolumn[
\hsize\textwidth\columnwidth\hsize\csname@twocolumnfalse\endcsname
\draft

\title{Effects of dissipation on quantum phase transitions}
\author{Aharon Kapitulnik$^1$, Nadya Mason$^1$, Steven A.
Kivelson$^2$, and Sudip
Chakravarty$^2$ }
\address{$^1$Departments of Applied Physics and of Physics, Stanford
University, Stanford, CA 94305, USA}
\address{$^2$Department of Astronomy and Physics, University of
California at Los Angeles, Los Angeles, CA 90095, USA}
\date{\today}
\maketitle

\begin{abstract}
We discuss the effect of dissipation on quantum phase transitions. In
particular we concentrate on the Superconductor to Insulator and 
Quantum-Hall to Insulator
transitions. By invoking a phenomenological parameter $\alpha$ to 
describe the coupling of the
system to a continuum of degrees of freedom representing the 
dissipative bath, we obtain new
phase diagrams for the quantum Hall and superconductor-insulator 
problems. Our main result is
that, in two-dimensions, the  metallic phases observed in finite 
magnetic fields (possibly also
strictly zero field) are adiabatically deformable from one to the 
other. This is plausible,
as there is no broken symmetry which differentiates them.
\end{abstract}

]

Quantum phase transitions continue to attract intense
theoretical and experimental interest; see, for example,
\cite{sondhi,elihu,chn,subir,haviland}. Such transitions -- where 
changing an external
parameter in the Hamiltonian induces a transition from one
quantum ground state to another, fundamentally different one -- have
been invoked to explain data from various experiments.
Transitions that have been studied include the
quantum-Hall liquid to insulator transition (QHIT), the quantum Hall liquid
to quantum Hall liquid or ``plateau'' transition (QHPT), the metal
to insulator transition (MIT) and the superconductor to
insulator transition (SIT). Where the transition is continuous, quantum
critical phenonema are expected to give rise to interesting, universal
physics which it is common practice to analyze using
a straightforward scaling theory, inherited from the classical theory
of finite temperature phase transitions.

Effects of dissipation, that is to say the coupling
of the critical modes to a continuum of other ``heat-bath'' degrees of
freedom, can fundamentally alter the character of the phases and
of the transitions between them.\cite{sudipandme,rimberg,takahide}
While in classical statistical mechanics, the dynamics and thermodynamics
are independent of each other, in the quantum case they are intimately
related.  The dynamical relaxation processes that permit the system to
reach equilibrium can be neglected in classical problems, but cannot
be  ignored in a quantum problem.

Recently, compelling experimental evidence has accumulated of the existence
of ``metallic'' phases, that is to say phases with finite dissipation in
the zero temperature limit.  There is as yet no  microscopic
understanding of these observations.  We {\it conjecture} that a metallic phase
is stabilized by strong-enough coupling to a dissipative  heat-bath,
which we characterize by a single phenomenological parameter,
$\alpha$, in a manner pioneered in early studies of macroscopic
quantum tunneling and coherence \cite{tony}.  Note that the present phenomenological
approach is impervious to such important issues as whether the dissipation is intrinsic
or extrinsic. For large enough $\alpha$, quantum coherence can be suppressed
even at zero temperature \cite{clastoquant}.  Thus, the conventional picture of
quantum phase transitions is certainly  dramatically altered, and 
``intermediate'' metallic phases can appear in the phase diagram.

\begin{figure}
\includegraphics[width=0.8 \columnwidth]{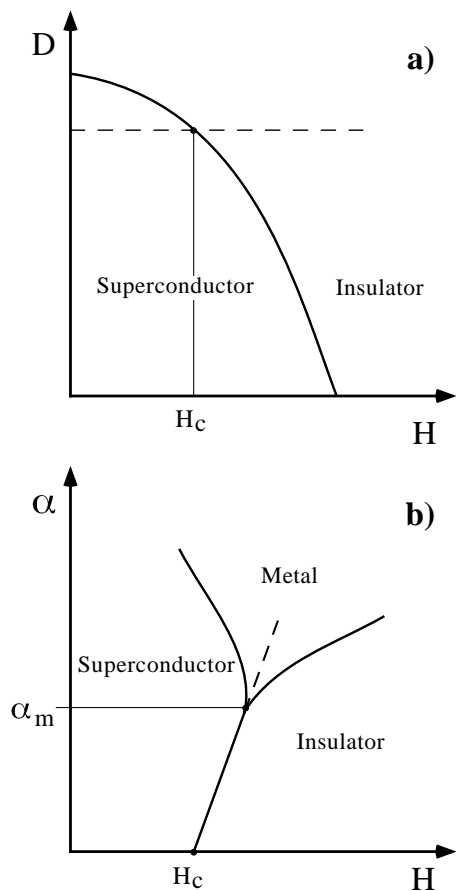}
\caption{Phase diagram for the field-tuned SIT: a) the H-D (D stands 
for Disorder) plane with the dashed line representing a plane at 
finite disorder, and b) H-$\alpha$ ($\alpha$ stands for Dissipation 
strength) diagram at finite disorder. $H_c$ marks the SIT critical 
point, and $\alpha_m$ marks the critical dissipation above which a 
metallic phase is obtained (see text for details.)
}
\label{fig1}
\end{figure}

As a paradigmatic example, consider the magnetic field driven SIT, for which
the commonly accepted phase diagram is shown in Fig. 1a. In the neighborhood
of the phase boundary, quantum critical scaling with universal exponents
is expected (so long as $D$ and $H\ne 0$);in particular the correlation length
exponent should be $\nu\approx 7/3$, as discussed below. With the introduction
of dissipation, the phase diagram is modified in a manner that is not 
presently well
understood.  Here, we {\it postulate} that, as originally proposed in Ref.
\cite{mason1}, the result is the phase diagram shown schematically in Fig.1b.
Here, when $\alpha > \alpha_m$, a finite resistance metallic state 
appears between the
superconducting and insulating phases. Moreover, to the extent that a 
crossover from a
positive to a negative coefficient of resistance occurs along the 
dashed line in the figure,
classical percolation, with $\nu=4/3$ will describe the physics at 
high temperatures, as is
indeed observed experimentally\cite{mason1}. However, to reconcile 
the fact that
finite dissipation in general tends to suppress quantum fluctuations 
and thus "pin" the
superconducting phase, the phase boundary emerging from $H_c$ is slightly
tilted towards high fields.

The same considerations have direct implications for quantum Hall systems.
Here, classical percolation in the limit of slowly varying
disorder\cite{trugman} and a crossover to quantum percolation near 
the transition
\cite{chalker,lee1} can be motivated from the microscopic theory.
It was previously shown \cite{klz}, by iteratively applying the
``Chern Simons flux attachment'' transformation \cite{zhk}, that a
global phase diagram for quantum Hall systems can be obtained from
considerations of the magnetic field driven SIT.  Here, the various
QHITs and QHPTs are mapped onto a single SIT. Consequently, if we adopt
the phase diagram in Fig.1a, the corresponding quantum Hall
phase diagram is that shown in Fig. 2a, with direct quantum
transitions between the various quantum Hall phases governed by
simple selection rules. However, the existence of the intermediate metallic
phase in Fig. 1b produces for the quantum Hall system the
new phase diagram shown in Fig. 2b, where for large enough $\alpha$,
each direct transition point opens up into a metallic regime.

This has many further consequences:

1)  In quantum Hall systems, in which metallic behavior is observed
below an apparent high temperature QHIT, behavior analogous to that
observed in the magnetic field driven SIT should be seen.  Among
other things, this means that in the high temperature scaling regime,
an apparent exponent $\nu=4/3$ should be observed, and that at fields
well larger or smaller than the apparent critical field, a true low
temperature metal insulator and metal to quantum Hall liquid
transition should be found.  Preliminary evidence of the correctness
of the first of these predictions is shown in Fig. 3.

2)  In quantum Hall systems, in which a direct QHIT is observed with
quantum exponents, $\nu\approx 7/3$, a transition to classical
percolation behavior with an intermediate phase can be induced by
increasing the dissipation in the system.  This can, in principal, be
done, using the strategy employed by Rimberg {\it et al.},\cite{rimberg},
by placing a second 2DEG which is capacitively coupled\cite{zimanyi} of the first 
and tuning the
conductivity of the second 2DEG by means of a back-gate.  The critical
phenomena associated with the value $\alpha=\alpha_m$ at which the
metallic phase first appears should be very interesting.

3)  Conversely, by reducing the amount of dissipation in the system
(for instance, by studying Josephson junction arrays or granular
films in a field) the intermediate metallic phase observed at low
temperatures in experiments on the field driven SIT should be
narrowed, and ultimately eliminated.  If this can be achieved,
rather than classical 
percolation exponents, values of $\nu\approx 7/3$ are expected.

\begin{figure}
\includegraphics[width=0.8 \columnwidth]{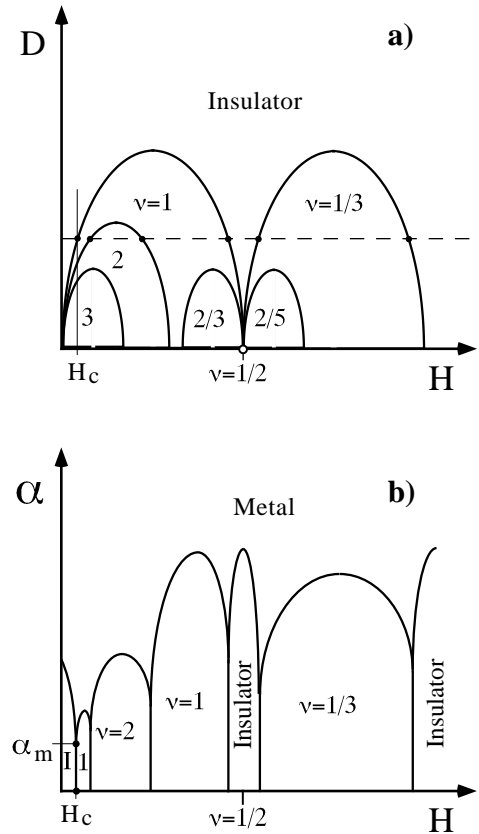}
\caption{Similar to figure 1 but for the quantum Hall problem.
a) The global phase diagram of Kivelson, Lee and Zhang {\protect\cite{klz}}
with dashed line representing a cut at finite disorder. Full circles 
represent the critical points obtained for a particular realization of disorder. 
b) H-$\alpha$ ($\alpha$ stands for Dissipation strength) diagram at 
finite disorder. This figure is  obtained by applying
Flux Attachment and the Law of Corresponding States (see section 
II-D), and it preserves the correct topology of Fig.1b (slopes of 
the phase boundaries at small $\alpha$ are omitted.)
$H_c$ marks the QHIT critical point from $\nu =1$ to insulator. Other 
critical points are found in a similar way.
}
\label{fig2}
\end{figure}

\begin{figure}
\includegraphics[width=0.8 \columnwidth]{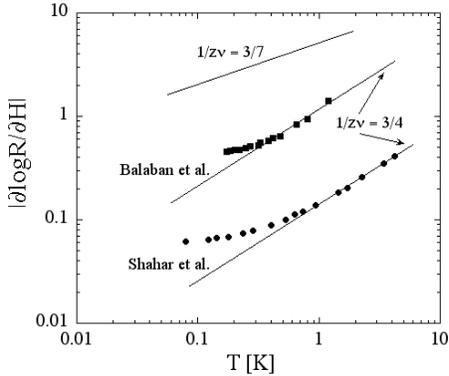}
\caption{Data taken from Shahar {\it et al} \protect\cite{shahar} and
Balaban {\it et al.} \protect\cite{balaban}. The straight lines with
slope $1/z\nu$=3/4 represent the asymptotic trend of the data at
"high temperatures" (see text for details.)  The straight line with a 
slope of $1/z\nu$=3/7
represents the expected behavior for a quantum percolation transition.
}
\label{fig3}
\end{figure}

4)  The flux attachment transformation to relate states at zero
$H$ to those at non-zero $H$ is subtle\cite{klz}.
Even if the average magnetic flux is canceled in this way, the
Hamiltonian at zero $H$ respects time-reversal symmetry at the microscopic
level whereas the finite field system does not.  However, if time 
reversal symmetry is
spontaneously broken even at $H=0$, as would occur for instance in a 
spin-glass phase, then
the correspondence between zero-field and finite $H$ phases might be
reliable.  In this way, a correspondence between the Hall metal and
the zero-field metallic phase might occur, with many
consequences\cite{sudipsideas}.

5)  All the metallic phases observed in these systems
at non-zero $H$ (and possibly even at $H=0$ if time
reversal symmetry is spontaneously broken) are adiabatically connected.

\section{Intermediate ``metallic'' phases}

The theory of a metallic phase in two dimensions at zero temperature
is a matter of intense
current debate\cite{mason1,sak,das,dalidovich,feigelman,spivak1,demler,lerner}.
Indeed, such a phase was thought for many years not to be possible
\cite{aalr,aal}. Recent  experiments have shown that metallic phases are
not only possible, but exceedingly common whenever interaction effects are
strong \cite{elihu}.  Of course, there is always a question whether the
metallic behavior is a finite temperature artifact, as there is no way to prove
that the resistivity would not diverge or vanish if the
temperature were lowered enough.  Below we list some of the most clear-cut
cases in which an apparently metallic phase has been observed.  Since in
each case, the experiments access temperatures low compared to all the simple
energy scales in the problem, we feel it is reasonable to
accept this evidence at face value.

1)  An apparently metallic phase  occurs in superconducting films in a magnetic
field for intermediate magnetic field
strengths \cite{mason1,hebard,paalanen,yazdani,ephron,valles} and in arrays of
Josephson-junctions \cite{zant}. In particular, this behavior is observed at
low temperatures in the magnetic field range at which, at higher temperatures,
scaling behavior is observed which was formerly interpreted 
\cite{mason1,ephron,mason2}
as indicative of a SIT.

2)  Analogous behavior has been observed in semiconductor heterojunctions for
magnetic fields in the neighborhood of a putative QHIT \cite{shahar,balaban}.
Specifically, metallic behavior is observed at low temperatures in 
the neighborhood
of the critical magnetic field at which higher temperature data is indicative
of a QHIT.

3)  In quench-condensed films of Ga, Pb and In at $H=0$, an anomalous 
metallic phase is
seen \cite{liu,dynes,goldman} below the local superconducting transition
temperature in which the resistance decreases strongly (by as much as 
five decades) with
decreasing temperature, roughly like $R=R_0 \exp[T/T_0]$, but 
extrapolates to a finite value,
$R_0$, as $T\rightarrow 0$.

4)  In Si MOSFETs, and other high mobility semiconductor
devices  which access the strongly interacting (large $r_{s}$) physics of the
two dimensional electron gas (2DEG), an unexpected metal insulator transition
at $H=0$ has now been widely and convincingly 
documented.\cite{elihu,kravchenko,gold}
The stability of the metallic phase has also been supported by
studies \cite{jiang12b,shaharnewun,simosfet}  of the quantum Hall 
effect at small non-zero
$H$:   Theoretically\cite{laughlin2,khmelnitskii}, a sequence of
transitions resulting from the ``floating up'' of the extended states 
is predicted
to occur as the expected insulating state is approached  as 
$H\rightarrow 0$, and
indeed this is observed in smaller $r_{s}$ devices \cite{wong,other}.
However, at large $r_{s}$, the delocalized states (or more precisely, 
the critical lines
separating different integer quantum Hall states) do not move up in 
energy as $H\rightarrow 0$,
thus allowing for a metallic state at $H = 0$.

5) A related set of experiments\cite{krav,song} on the behavior of high
mobility 2DEG's in the small magnetic field limit show behavior which
we interpret as indicative of a metallic phase for a range of
weak magnetic fields. Note, these experiments were interpreted
somewhat differently by their authors:  The data clearly reveal a 
failure of the
delocalized states to ``float up'' as $H\rightarrow 0$ and a 
breakdown of the selection
rules\cite{klz} thought theoretically to govern quantum Hall 
transitions, including
the QHIT.  However, the measurements were originally interpreted in 
terms of a QHIT
in which the ground-state at $H$ smaller than a (electron density dependent)
critical field, $H_{c}$ was said to be insulating. Indeed, in this 
field range, the
resistance increases with decreasing  temperature, but only weakly, 
and it appears to saturate
at low temperatures.  In the following, we will re-interpret this 
data in terms of a quantum
Hall liquid to metal transition, rather than as a ``new universality 
class of QHIT.''

6)  Metallic phases were also found, under special circumstances,
in high quality devices (especially in GaAs heteronjunctions)
in the presence of a ``quantizing''  magnetic field ( {\it i.e.} $H$
big enough that the cyclotron energy, $\hbar\omega_{c}$, is larger 
than the temperature,
the disorder potential, and even the Coulomb strength).   This ``Hall 
metal'' has been
observed \cite{jiang12} and extensively studied \cite{half,nu12} for 
a  {\it range}
of magnetic fields near ``filling factor'' $\nu$=1/2 ({\it i.e.} two 
magnetic flux
quanta per electron) and related even fractions.  Precisely at 
$\nu=1/2$ there is some
indication of an upturn in the resistance at the lowest 
temperatures,\cite{vgoldman,wong2},
which might indicate that the true ground-state is insulating, but 
overall in this range of
fields, the proponderance of experimental evidence supports the 
existence of  a true metallic
phase.

7)  In weakly disordered metallic films (with large values of 
$k_F\ell$) the phase
coherence length, which according to theory controls the low 
temperature behavior of ``weakly
localized'' systems, appears \cite{web} to saturate at low 
temperatures (The determination of
$\ell_\phi$ was done using magnetoresistance measurements,) rather 
than diverging as required by
theory \cite{phasediverges}.

\section{Extremal zero temperature phases}

In this section we summarize our theoretical understanding of the 
relevant zero temperature
electronic phases in two spatial dimensions in the absence of 
dissipation ($\alpha=0$).
This discussion is not meant to be exhaustive as there are numerous 
other phases, for example
phases that spontaneously break time reversal 
symmetry\cite{sudipsideas,laughlin}, which may be
important under some circumstances.

\subsection{The superconducting state}

Numerous two dimensional systems in zero magnetic field exhibit a
superconducting or superfluid phase below a non-zero transition
temperature. The theoretical understanding \cite{superconducting} of this
state is not in question, as far as we know: at finite temperature, 
the superfluid
phase is characterized by a non-zero superfluid
density, but no broken symmetry, although in the zero temperature limit,
a true broken symmetry state is expected.
The zero temperature superconducting state persists in the presence of
a non-zero magnetic field, $H$, so long as it is small enough,
$H < H_c$.  In the absence of disorder, this is due to the crystallization
of the field induced vortices and pinning at the boundaries (note that with no
pinning at the boundaries, any finite current will cause the vortex 
lattice to slide and
thus result in dissipation);  this vortex crystal (Abrikosov lattice), itself,
persists to a finite melting temperature, so in this limit 
superconductivity survives at
low, but non-zero temperatures.  In the presence  of quenched 
disorder, the vortex
crystal at small $H$ is certainly disrupted
\cite{braggglass}, but at $T=0$ the dilute vortices are
localized by the disorder.  However, in this case superconductivity
is destroyed at any non-zero temperature due to thermally
activated vortex motion \cite{fisher1}.  Thus, in systems of
interest, for which disorder is an important feature of the physics, no  $T>0$
transition to the superconducting state is expected;  in
this phase the conductivity diverges continuously as $T\rightarrow 
0$, together with a
diverging length that describes the proximity to this 
zero-temperature vortex glass
phase. Consequently the $I-V$ characteristics of the system will 
exhibit non-linear
behavior above some threshold current which itself vanishes as the
temperature tends to zero \cite{hyman}.

\subsection{The insulating phase}

In the limit of large disorder, large $H$, or strong interactions 
between particles,  all the
particles are localized and an insulating ground state occurs.  This 
is independent of whether
the constituent particles are taken to be bosons or fermions (or 
anyons, for that
matter). In the absence of disorder, this insulating state has Wigner
crystalline long-range order, and correspondingly a finite 
temperature phase transition.
However, disorder should couple to the charge order of the Wigner 
crystal as a random
field. Thus, from the Imry-Ma random field arguments, no true finite
temperature phase transition to the insulating  state occurs.  The insulating
state is then characterized by a resistivity which diverges continuously as
$T\rightarrow 0$.

In the absence of a magnetic field, a further distinction may, in 
principle, distinguish
various insulating phases based on their spin (magnetic) structure
\cite{philmag,randomsinglet}. Electrons in the extreme low
density limit in the absence of disorder form a Wigner crystalline
state in which the electron spins are ferromagnetically ordered 
\cite{ferro}.  Although even
weak disorder eliminates long-ranged Wigner crystalline order, 
presumably the ferromagnetism
would survive up to a critical disorder strength.
At higher densities, but still in the Wigner crystalline state, 
antiferromagnetic
exchange processes begin to dominate, so a frustrated antiferromagnet, possibly
with some form of spin-pairing and a spin gap, may occur 
\cite{philmag,luhlier,rvb}.
Such an insulating state with a spin-gap could also occur as the 
result of a localization
transition of a system of Cooper pairs, as would be expected, for 
instance, in the Coulomb
blockade limit of a granular superconductor \cite{pairedinsulator}.
If so, some form of spin pseudo-gap is likely to survive the introduction of 
weak disorder.
At large disorder, a zero temperature spin-glas 
s phase is also conceivable.
At non-zero Zeeman coupling to an external magnetic field, most or all of the
magnetic distinctions between various possible insulating phases are removed.

A distinction between insulating states in a magnetic field has also 
been discussed
based\cite{orderoflimits} on the asymptotic behavior of the Hall 
resistance, $\rho_{xy}$, as
$T\rightarrow 0$.  A Wigner crystal, and presumably also an Anderson 
insulator\cite{imry} have
a Hall resistance which diverges in this limit.  Conversely, it has 
been proposed\cite{klz} that
a ``Hall insulator'' phase exists which has a vanishing conductivity 
tensor, but a Hall
resistance that approaches a finite value, presumably of order $H/nec$, as
$T\rightarrow 0$.  A still more exotic, ``Quantized Hall Insulator''
phase has also been proposed\cite{quantizedinsul} in which, despite the
vanishing of the conductivity tensor, the Hall resistance approaches 
a quantized
value, for instance $h/e$, as $T\rightarrow 0$.  It is likely, 
although not proven, that a Hall
insulator is a distinct state of matter, and must be separated by a
phase transition from an ordinary insulator.  This is certainly
the case for a quantized Hall insulator,\cite{klz} as there must be
a first magnetic field at which the Hall resistance ceases to
be quantized.

In the present paper, we will always assume that the effects of disorder
are important, so we will not explicitly discuss the Wigner crystal or
Abrikosov lattice.  For simplicity, we will also neglect the various
magnetic and transport distinctions between insulating phases, as well.

\subsection{The quantum Hall liquid}

The quantum Hall liquid has a vanishing longitudinal conductivity and 
a quantized Hall
conductance, $\sigma_{xy}=(e^2/h) s_{xy}$ where $s_{xy}$ is an 
integer or one of a particular
set of rational fractions for the the integer and fractional Hall 
effect, respectively.
There is no true broken symmetry, and so no finite temperature phase transition
to this state.   Quantum Hall liquid states are partially characterized by the
quantized value of the Hall resistance. A quantum Hall state is
typically most stable at a ``magic'' value of the magnetic field,
$H_{magic}=s_{xy}^{-1}\phi_0\bar\rho$, where $\phi_0=hc/e$ is the
quantum of flux and $\bar\rho$ is the mean electron density.  However, like
the superconducting state, the quantum Hall state at $T=0$ is stable 
for a finite
range  of disorder and magnetic field.

Quasi-particles are generated in the ground-state when the magnetic field
differs from the magic value, or can be nucleated by disorder.  They
play a role analogous to that of vortices in a superconductor 
\cite{zhk}, and when
they are not localized they destroy the quantum Hall state.  In the absence
of disorder, a low density of quasi-particles will crystallize,
producing a quantum Hall state which is analogous to the Abrikosov lattice.
In the presence of disorder, the dilute quasi-particles are localized.
However, whereas the vortices in a superconductor are bosonic 
\cite{fisherlee}, the
quasi-particles in a quantum Hall state have statistics, fermionic in 
the integer
case and anyonic in the fractional case, which are 
related to $s_{xy}$.

\subsection{Flux attachment and the Law of Corresponding States}

There is a formal transformation \cite{zhk}, often called flux
attachment, which maps a system of interacting particles in a magnetic field
with Bose or Fermi statistics to another problem of {\it transformed}
``Chern-Simons'' particles with the same mass, same interactions, but 
with modified statistics,
and which interact with a fluctuating ``statistical gauge field,'' 
$a_{\mu}$, whose dynamics
are governed by a Chern-Simons term, rather than the usual Maxwell action.
The Chern-Simons action has the effect of attaching $\theta$ (related to the
Chern-Simons coupling constant) quanta of statistical flux to each particle.
Consequently, the {\it net} magnetic flux seen by the transformed 
particles  is modified
according to

\be
H^{eff} = H - \theta \phi_0 \bar\rho,
\ee

and the phase (statistical) angle
associated with the interchange of two particles is transformed according to

\be
\phi_{exchange}=(n - \theta)\pi
\ee

where, depending on whether the original particles are
bosons or fermions, $n=0$ or $n=1$, respectively.  Manifestly, for 
$\theta$ an odd integer,
this transformation maps bosons into fermions and vice versa, while 
for $\theta$ even, it maps
bosons into transformed bosons, and fermions into transformed fermions.

The effect of this transformation is a formal relation between
seemingly different states of matter.  For instance, if for some 
choice of $\theta$, the
resulting Chern-Simons bosons are condensed into a superconducting 
state, the original particles
are necessarily in a quantum Hall liquid state, with conductivity tensor

\be
\sigma_{xy}=(q^2/h) (1/\theta), \ \ {\rm and} \ \sigma_{xx}=0,
\label{eq:sxy}
\ee

where $q$ is the particle charge ({\it i.e.} the coupling to the external
electromagnetic gauge field).  Conversely, if the Chern-Simons bosons are
``localized'' in an insulating state, $\sigma_{ij}=0$.
In similar ways, it is possible to relate fermions in an integer 
quantum Hall state to
electrons in a fractional quantum Hall state\cite{jain,lopezfradkin}, etc.

There is another transformation, which is a version\cite{fisherlee}
of a standard duality transformation, that relates the physics of 
different phases,
and which can be implemented\cite{klz,qhreview} straightforwardly in 
the bosonic
Chern-Simons field theory.  Formally, this transformation relates the 
particle three
current, $J_{\mu}$ to a dual gauge field, $a_{\mu}^{dual}$,

\be
J_{\mu}=\epsilon_{\mu\nu\lambda}\partial_{\nu}a_{\lambda}^{dual}
\ee

and conversely,  a dual particle (vortex) current to the original 
gauge field, $a_{\mu}$,

\be
J_{\mu}^{dual}=\epsilon_{\mu\nu\lambda}\partial_{\nu}a_{\lambda}
\ee

where $\epsilon_{\mu,\nu,\lambda}$ is the Levi-Cevita symbol.  The remarkable
feature of this transformation in 2+1 dimensions is that the action
has the same form in terms of the original and transformed variables 
but the statistical angle
is transformed according to

\be
\theta^{dual}=-1/\theta.
\ee

Moreover, when considering the coupling to an external
electromagnetic gauge field the particle charge, $q$, transforms according to

\be
q^{dual}=-q/\theta
\ee

and the conductivity tensor (computed from the Kubo formula) for the original
variables is related to that computed in the dual theory according to

\be
\sigma_{ij}=\sigma_{ij}^{dual} + (q^2/h)(1/\theta)\epsilon_{i,j}
\label{eq:sxydual}
\ee

(This expression appears to differ from, but is actually consistent with a more
familiar relation which involves the one gauge-field irreducible
response functions of the Chern-Simons particles, {\it i.e.} the 
particle response to the
combined external and statistical gauge fields.  In terms of these 
irreducible response
functions, the conductivity  of the original particles is related to 
the resistivity
of the dual particles.)

Comparing the expressions in Eq. (\ref{eq:sxy}) and Eq. (\ref{eq:sxydual}),
it is clear that if a set of Chern-Simons bosons are condensed in a
superconducting state, the dual bosons must be localized, and 
conversely.  In the
context of the SIT, this allows one to view the insulating state as a 
condensed state of
vortices\cite{fisherlee}.  In the case of the integer quantum Hall effect with
$s_{xy}=1$ (and hence $q=e$ and $\theta=1$), the quantum Hall liquid state can
be viewed as a condensed state of charge $e$ bosons, with vortex excitations
(quasi-holes) with charge $q^{dual}=-e$, fermionic statistics
($\theta^{dual}=-1$), while the proximate insulating state can be viewed as a
vortex condensate, and the original electrons become the vortices in 
the dual theory.

These two transformations form the basis for a
Law of Corresponding States,\cite{klz,jtk} which relates seemingly
different states of matter. However, it should be recalled that, because the
fluctuations of the statistical gauge field induce additional 
interactions between
particles, at a microscopic level the correspondence may be quite complicated.
However, if only the topology of the phase diagram and the nature of 
the phases are
of principal interest, the Law of Corresponding States can be adopted 
without caveat.

Within the context of the dirty boson model, it is thought that there are only
two zero temperature phases\cite{caveatboson} in the presence of disorder, a
superconducting phase and an insulating phase, as shown in figure 1a.
Assuming that this is the case for the Chern-Simons bosons in the 
quantum Hall effect,
a global phase diagram for the quantum Hall systems can be 
constructed.\cite{klz}
The result is summarized in figure 2a.

\section{Critical phenomena}

\subsection{Quantum phase transitions}

It is clear that an understanding of the phases is a necessary
precursor to an understanding of the quantum critical phenomena associated with
the transitions between them.  However, unlike the analogous classical problem,
it is possible for the universality class of a quantum transition to depend on
the nature of the dynamics as well as on the character of the two phases.

The simplest possibility, about which the most is known theoretically, is that
some or all of the observed transitions are equivalent to a SIT in a system of
disordered, interacting fundamental bosons, the so-called ``dirty
boson'' problem. Because the electrons form Cooper pairs in the 
superconducting state,
and because it is ultimately fluctuations in the phase of the order parameter
which are expected to destroy superfluidity (so that the electrons 
can plausibly be
viewed as paired even on the disordered side of the transition), it 
is an appealing view that
the dirty boson problem captures the universal features of the SIT in actual
superconducting films.  One of the most striking predictions that has
been made on the basis of this model\cite{fisher2} is that the 
conductivity tensor at
the critical point takes on a universal value in units of $(e^*)^2/h$ 
(where $e^*=2e$
for Cooper pairs and $e^*=e$ for electrons in the quantum Hall 
effect).  To the best of our
knowledge, there are no convincing calculations, either numerical or 
analytic, for the
value of the critical conductance or the critical exponents at this
transition.  An argument can be made on the basis of particle-hole
symmetry that the Hall conductance at the transition is 0. 
Independent arguments,
based on the $1/r$ form of the Coulomb interaction, lead to the 
expectation that
the dynamic exponent $z=1$.

The most obvious, potentially dangerous piece of physics
that is omitted in this approach is the effect of low energy
quasi-particle excitations in such systems.  Presumably, in a 
granular superconductor,
where a relatively clean, large gap in the quasi-particle spectrum is 
expected and,
indeed, observed \cite{dynes,hsu}, the neglect of quasi-particles is 
a safe bet.  However, to
date, nothing resembling the expected SIT has been observed in 
granular films - what is observed
is best described as a crossover (an actual transition point is 
difficult to identify in the
data) from a superconductor to strange metal to an insulator.  In 
disordered films, even in the
absence of a magnetic field, one would expect on theoretical grounds 
that in the neighborhood of
the critical film thickness, the superconducting gap will be filled 
in with a large density of
gapless quasi-particle excitations, and this expectation is 
apparently born out in experiment.
In the presence of a magnetic field, where gapless quasi-particle 
excitations are expected in
the cores of vortices, this expectation is even stronger,
especially when the spacing between vortices is not large compared to 
their radius.  In both
cases dissipation due to gapless quasi-particle excitations can
potentially alter the critical phenomena.

In quantum Hall systems a variety of zero temperature continuous
phase transitions are expected and, in one form or other, observed: 
1)  The QHIT, both for the
case in which the quantum Hall liquid is an integer Hall state (typically, 
$s_{xy}=1$ or, when
spin-splitting are not resolved, $s_{xy} 
=2$) or a fractional Hall state (typically
$s_{xy}=1/3$). 2) Transitions (``plateau transitions'') between two 
distinct quantum Hall liquid
states, which again might be integer or fractional. For non-interacting
electrons, the critical exponents associated with the QHIT are known
to be $\nu\approx 7/3$, and an argument due to Lee and Wang 
\cite{leewang} shows that at least
short-range interactions do not alter this result.  Under conditions 
of particle-hole
symmetry, it can be proven\cite{klk} quite generally that the 
critical Hall conductance at
$s_{xy}=1$ quantum Hall liquid to insulator transition is 
$\sigma_{xy}^{c}=(e^2/h) (1/2)$;  if
the critical conductance is, indeed, universal then this must be the 
critical value, even in the
absence of (microscopic) particle-hole symmetry. An argument based on 
particle-vortex duality in
the Chern-Simons formulation of the problem leads (among others) to 
the relation (also
called\cite{ruzin,sondhi2,klz} the ``semi-circle law'') that
$\rho_{xx}=(h/e^2)$ in the vicinity of the $s_{xy}=1$ or $s_{xy}=1/3$
to insulator transition.  Together, these two results lead to the 
conclusion that
$\sigma_{xx}^c=\sigma_{xy}^c=(e^2/h) (1/2)$ at the $s_{xy}=1$ to 
insulator critical point, a
conclusion that has also been supported by numerical studies of 
non-interacting electrons
\cite{bhatt,lee1}.

At least superficially, there is less reason to necessarily expect
additional dissipation in quantum Hall systems than in 
superconducting films.  However, in
addition to the many possible extrinsic sources of dissipation, there 
are certainly ways such
dissipation could arise intrinsically, even here.  For instance, there could be
\cite{leeprivate,chlovskii} compressible regions of electronic 
structure  associated with
the ``Hall metallic'' or $\nu=1/2$ state.  These might arise 
naturally, especially in systems
with a smoothly varying disorder potential, in the form of ``fat'' edge states.

A quantum model for dissipation consists of coupling the system
variables to a continuum of Ohmic heat bath degrees of freedom; dissipation
is then merely a flow of energy from the system to the heat 
bath\cite{tony}. The
integration of the heat bath degrees of freedom results in an action
containing induced temporal long range interactions. The analysis of such
long range interactions brings in a new control parameter signifying the
coupling to the dissipative heat bath, $\alpha$, proportional to 
$h/e^2 R$, where $R$ is
the shunt resistance characteristic of the dissipative heat bath. In
addition the action will depend on the dimensionality of the space, 
the symmetry of the  order
parameter, etc., common in a conventional description of a quantum phase
transition\cite{sondhi,subir}.

A quantum phase transition often involves two related phenomena: the
formation of a condensate and the associated  metastability. For a
superconductor or a superfluid, this is widely appreciated. In order to
stabilize the superconducting state the phase slip processes must be suppressed
and the vortices must be localized. In an incompressible quantum Hall 
state there is
no true broken symmetry, but its metastability hinges upon localizing the
quasiparticles over a finite range of disorder and magnetic field (the
fermionic quasiparticles playing the role of vortices in a superconductor.)
Thus, in the limit of large dissipation, a quantum phase transition 
can acquire a very
special character because dissipation can efficiently reduce quantum
fluctuations of the system. As a result, the competition between the 
kinetic and potential
energies is altered, resulting in a state that is characteristically classical.
In other words, the formation of the condensate is tuned by dissipation.
Similarly, the metastability of the state can be altered by suppressing
quantum tunneling and motions of vortices and quasiparticles, typically by the
orthogonality overlap of the continuum degrees of freedom of the heat bath.
Such a dissipation tuned transition is well documented in the
case of a single Josephson junction, in which as a function
of $\alpha$, the system undergoes a transition from a
quantum state at $\alpha<\alpha_c$, where quantum diffusion of the 
phase destroys the
superconducting state, to a  ``classical'' state at 
$\alpha>\alpha_c$, where quantum tunneling
is  suppressed and  the junction is truly superconducting at 
$T\rightarrow 0$.  Such a
transition  can occur in higher dimensional systems, as well, as has 
indeed been indicated in
recent experiments on Josephson Junction arrays \cite{rimberg}.

Because the first effect of dissipation is to suppress fluctuation, and
hence to stabilize the superconducting phase, we have drawn the 
superconducting to
insulating phase boundary, $H_c(\alpha)$ in Fig. 1b with a positive 
slope. However, in the limit
of large $\alpha > \alpha_{m}$, we have indicated a metallic phase, 
as is suggested, for
instance, by the recent work in Ref. [24]. The reason for the 
existence of the metallic phase is
that with the suppression of quantum fluctuations, the motion of the
excitations responding to the external field is diffusive and 
classical, with  characteristics
inherited from the heat bath.  It is likely that particle exchange 
with the heat
bath, in addition to the usual capacitive coupling, will play an 
important role in this
physics.  The existence and the stability of a metallic phase at 
large $\alpha$ is the central
postulate of the present work whose theoretical basis is presently uncertain.

\subsection{Classical percolation}

There is another limiting view of any transition between a
conducting (or superconducting) and an insulating phase provided by 
classical percolation;  as
the conducting regions of the system percolate, the system goes from 
being globally insulating
to globally conducting.  In two dimensions, many of the critical 
properties are well understood
theoretically, including the correlation length exponent, 
$\nu_{perc}=4/3$.  The correlation
length $\xi_{perc}$ is, roughly speaking, the radius of gyration of 
the largest typical clusters
of the minority phase \cite{as}.  In the case in which the length scale of the
disorder is very long, so that quantum tunneling, and more 
particularly quantum coherence
between distinct tunneling events can be ignored, both the QHIT and 
the SIT would
be expected to be well approximated as percolation transitions.  The 
earliest theories of the
QHIT were based on percolation of puddles of Hall liquid 
\cite{trugman}.  Likewise, the
phenomenologically successful puddle theory of Shimshoni, Auerbach, 
and Kapitulik \cite{sak} is
based on the assumption that the measured transport is dominated by 
tunneling of
vortices (in the case of the SIT) or quantum Hall quasi-particles
(for the QHIT) across a characteristic, isolated weak-link between puddles.

It is generally believed that at low enough temperature, any
classical percolation
approach must break down.  Roughly speaking, if we imagine that,
as in weak-localization theory, there exists some sort of quantum 
phase coherence length,
$\ell_{\varphi}(T)$ that diverges as the temperature tends to zero, 
then the classical theory
would be valid so long as $\xi_{perc} \gg \ell_{\varphi}$, with the 
true quantum behavior
apparent only at such low temperatures that $\xi_{perc} \ll 
\ell_{\varphi}$.  In the QHIT, this
crossover from classical to quantum percolation has been studied 
explicitly \cite{leewang}.

In figure 4, we illustrate this crossover schematically for the 
simplest case in which classical
percolation and the quantum phase transition occur at the same 
critical field, $H_c$. (In the
case of the $s_{xy}=1$ QHIT, particle-hole symmetry in the lowest Landau level
insures that this is the case.) Under the assumption that 
$\ell_{\varphi}$ diverges as
$T\rightarrow 0$, the $x$ axis is essentially the temperature. 
(Specifically, if
$\ell_{\varphi}(T)\sim T^{-1/p}$, then $x\sim T^{4/3p}$.) The dashed 
line in the figure denotes
$\ell_{\varphi}(T)= \xi_{perc}$, while the solid line denotes 
$\ell_{\varphi}(T)= \xi(H)$.
In the lightly shaded regime labeled ``classical percolation,'' where
$\ell_{\varphi} > \xi_{perc}$, the system can be viewed as  a 
collection of effectively
macroscopic droplets of the two-phases, and hence some sort of 
quasi-classical puddle theory
applies. In the darker shaded regime, between the dashed line and the 
solid line, at
which , the system is in the quantum critical regime, where 
$\xi(H)\sim |H-H_c|^{-\nu}$.  In
ordering these two crossovers, we have assumed that $\nu > 
\nu_{perc}=4/3$, which is certainly
satisfied if, as discussed, $\nu\approx 7/3$. For very large $\ell_{\varphi}$,
as T approaches zero, we allow for classical percolation to cease 
being relevant, hence the
darker shaded area covers all the region between the solid lines near 
$H_c$.  Below the solid
line the properties are dominated by dilute thermal excitations above the
appropriate ground-state phase, a regime that in other contexts 
\cite{chn,subir} is
called ``renormalized classical.''

Note that figure 4 describes a situation in which there is only one 
critical field for both,
classical and quantum percolation. For self dual systems in two dimensions the
percolation threshold has to be 50$\%$ for either case. However, for 
general systems the two
critical fields will be different. Indeed, where the puddles of the 
two competing
phases are not macroscopic ({\it i.e.} when the correlation length for
the disorder potential is finite) the notion of classical percolation
is not completely well defined, as there is an intrinsic ``quantum
blurriness'' to the edges of the puddles, so there is no precise point
at which two 
puddles can be said to touch.

\begin{figure}
\includegraphics[width=0.8 \columnwidth]{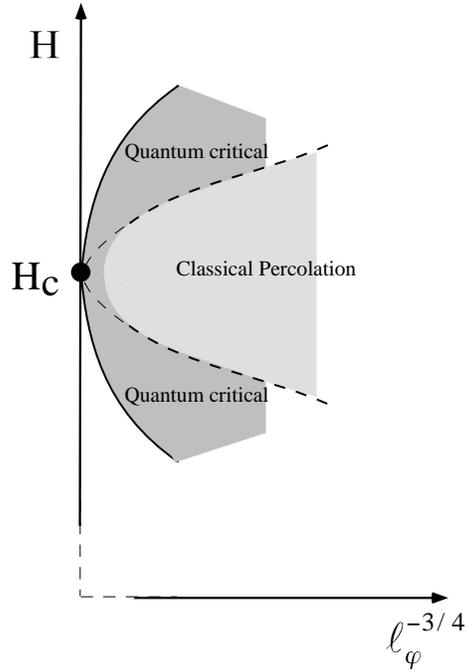}
\caption{H-T diagram near the SIT quantum critical point. Here we assume that
the "classical" and quantum critical points coincide. Inside the dashed lines
and in the lightly shaded area dissipation is important and the 
system is classical.
However, outside the range of influence of the classical point and 
between the two
solid lines the system (dark shaded area) is quantum critical. Also, 
as $\ell_{\varphi}$
diverges ({\it e.g.} at low enough temperatures), and with a 
diminishing influence of
dissipation, the system becomes quantum critical as well (see text.)
}
\label{fig4}
\end{figure}

However, {\em if} for some reason, perhaps due to strong enough 
coupling to a heat bath,
$\ell_{\varphi}$ does not diverge\cite{web} at low temperatures, then 
near enough to the
percolation threshold a sort of ``metallic'' phase, in which the dissipation
occurs \cite{sak} predominantly at weak-links between nearly 
percolating puddles
of the minority phase, will be valid to arbitrarily low temperatures!
Moreover, in this case, a true phase transition separates the 
metallic state in the
neighborhood of $H=H_c$ and the extremal quantum states at large 
values of $|H-H_c|$.
At present, we know of no theoretically well understood prototypical
systems in which this sort of behavior occurs.  Some very promising
starts along this line, especially related to the behavior of
superconducting grains in a metallic host, are enumerated in the final
section.  Nonetheless, it is important to realize that analogous behavior is
established in certain well understood zero dimensional
systems \cite{clastoquant}. Here, as a function of the strength of 
coupling to a heat-bath,
various quantum systems with more than one classical ground-state are
observed to make a transition from the expected quantum behavior, in
which tunneling renders the ground-state unique, to zero
temperature ``classical'' behavior, in which the system is trapped in
a single (arbitrary) one of the classical ground states.

\subsection{``Superuniversality''}
\label{sec:super}

The Chern-Simons mapping suggests\cite{klz,jtk} that there should be
a correspondence not only between the various stable electronic 
phases in two dimensions, but
also between the quantum phase transitions between them.  This idea 
is referred to as
``superuniversality''\cite{lutken}. Superuniversality as an 
approximate statement\cite{klz}
follows from treating the fluctuations  of the statistical gauge 
field at linear response
(one-loop) level. Non-perturbative proofs of superuniversality
have been constructed for certain simplified models, without 
disorder\cite{universality}.
Whether it is truly an exact relation asymptotically deep in the 
critical regime, or whether the
effects of higher order fluctuations of the statistical gauge field 
ultimate destroy
superuniversality, is still being debated.\cite{nonuniversal} If we 
assume that the
correspondences implied by suepruniversality hold to a sufficient level of
approximation, we can derive a large number of additional results.

The critical exponents, $\nu$ and $z$, should be the same at all quantum Hall
transitions, including the QHIT and the plateau transitions, 
irrespective of whether the
transitions involve integer or fractional quantum Hall states. 
Assuming that the
critical conductivity tensor is indeed universal, one can compute the 
critical conductance at
any given transition from its value at the $s_{xy}=1$ to insulator 
transition, as discussed in
Ref. \cite{klz}.

The same correspondence implies that the magnetic field driven SIT 
should have the same critical
exponents and a related value of the critical conductance.  In 
particular, if we
accept that at the $s_{xy}=1$ to insulator transition, 
$\sigma_{xy}^c=\sigma_{xx}^c=e^2/2h$,
then at the SIT we would expect $\sigma_{xy}^c=0$ and $\sigma_{xx}^c=2e^2/h$.

\section{Discussion}

While the classical theory of phase transitions has been
extraordinarily successful,
there are several
reasons to exercise caution when applying this approach to zero temperature,
quantum phase transitions:  1) Experiments are always
carried out at finite temperature, so that the proper identification of
the relevant phases requires an extrapolation to zero temperature.
In all the cases cited above, the question has arisen whether a 
transition between two
distinct zero temperature phases has actually been observed, or 
whether a finite temperature
crossover behavior is being misinterpreted as a quantum phase transition.
2)  In most experimentally interesting cases, quenched disorder is known to
play a central role in the critical physics, and there is increasingly
compelling theoretical reasons\cite{danfisher} to believe that 
Griffiths singularities,
which are typically only of academic interest for classical critical 
phenomena,  can
fundamentally complicate the scaling analysis at quantum critical 
points.  In particular they
can, under some circumstances, lead to divergent susceptibilities and 
relaxation
times over a finite range of parameters about the quantum critical point, and
apparent correlation length exponents which depend on the averaging
procedure.   3)  Effects of dissipation, that is to say the coupling
of the critical modes to a continuum of other ``heat-bath'' degrees of
freedom, can fundamentally alter the character of the phases and
of the transitions between them.\cite{sudipandme,rimberg}

{\it The SIT:}
On the basis of the dirty-boson model, the zero temperature phase diagram of
superconducting films is expected to be of the form shown in figure 1a;
all along the phase boundary, the SIT is in a single universality 
class, except at the zero $H$
end point.

How well are these expectations met experimentally?
Certainly, at not too low temperatures, a field
driven SIT has been apparently observed in a number of
superconducting thin-film systems.  Moreover, the
resistivity (more precisely the $I-V$ curve) as a function of the
$H$, $T$, and $I$ can be successfully collapsed onto a scaling curve, 
suggestive of quantum
critical scaling.  However, while the value of $z\approx 1$ extracted 
from the data is
consistent with theoretical expectation, it is always found that 
$\nu\approx 4/3$, suggestive
more of classical than of quantum percolation. Moreover, the critical 
resistivity is not found
to be universal, and is often as much as a factor 10
smaller than the value $\sigma_{xx}^c=2e^2/h$, predicted on the basis 
of superuniversality.
Finally, and most dramatically, it has been emphasized in the recent 
work of Mason
and Kapitulnik (although it is a highly neglected fact that is
lurking in much previous data, as well) that at low temperatures, the
resistivity saturates to a strongly field dependent but temperature 
independent value.  {\it
Thus, in fact no SIT actually takes place in this field range! } 
Manifestly, there is physics
here that is beyond the scope of the dirty boson model.

This observation does not contradict the theoretically solid
expectation that superconductivity will occur for a non-zero range of 
$H$.  Indeed, Mason and
Kapitulnik \cite{mason2} have recently found evidence
of a low temperature Metal-to-Superconductor transition in disordered
amorphous films in which a field-tuned Superconductor-Insulator 
Transition is disrupted
\cite{mason1}.  This new transition is characterized by hysteretic 
magnetoresistance and
discontinuities in the $I-V$ curves. The metallic phase just above 
the transition is different
from the "Fermi Metal" before superconductivity sets in as is evident from the
temperature dependence of the resistance and the $I-V$ 
characteristics obtained for this phase.
Of course, experiments at much higher $H$ to look for the metal to 
insulator transition
predicted by figure 1b are the next step.

{\it The QHIT and the QHPT:}
The Law of Corresponding States applied to the simple phase diagram 
in figure 1a leads to
the complex phase diagram in figure 2a, and superuniversality would 
imply that all
the phase transitions are in the same universality class.

How well are these expectations met experimentally?
The answer is mixed.  There is a subset of experiments that are in 
striking agreement
with these expectations.  Early experiments on the plateau 
transitions between various integer
quantum quantum Hall states, and from the $s_{xy}=2/5$ to 
$s_{xy}=1/3$ quantum Hall states
exhibited good scaling properties characteristic of a quantum phase
transition, and apparently universal values of the critical exponent, 
$\nu z \approx
7/3$.  Measurements at finite frequency showed remarkable $\hbar 
\omega/k_B T$ scaling, again
strongly indicative of quantum critical behavior.  No sign of 
saturation of the critical
behavior was detected to the lowest temperatures.  However, no clear results
concerning the critical conductance were obtained in these experiments.

Somewhat later, experiments were carried out\cite{jiangqhit} on
the QHIT, of which some of the clearest made use of purposely low
mobility heterojunctions.  Transitions from the $s_{xy}=2$ to
insulator (the factor of 2 is due to the fact that the spin-splitting is not
resolved in these samples) and the $s_{xy}=1/3$ to insulator 
transition.  Quantum critical
scaling is observed on some samples \cite{trivedi}, with $\nu 
z\approx 7/3$. Moreover, the
theoretically predicted universal values of the critical resistance 
are found to good
approximation. (Ambiguities due to the exact geometry of the current 
paths make a precise test
of this prediction difficult.) Some features of the shape of the 
global phase diagram in figure
2a were also confirmed in these experiments, including the existence 
of a reentrant insulator to
quantum Hall liquid to insulator transition as a function of 
increasing $H$ (also known as the
phenomenon of ``floating'' of the delocalized states 
\cite{laughlin2,khmelnitskii}).  However,
other experiments point to the phase diagram of figure 2b by showing 
deviations from scaling on
both sides of the transition \cite{hilke}. It is important to note 
that deviations from scaling
can occur even without noticeable resistance saturation, and it is 
therefore difficult to
determine the conductance states in experiments where scaling is not measured.

A large number of more recent experiments on the QHIT, carried out on 
high mobility
heterojunctions, exhibit behavior that is more like that observed in 
superconducting films: An
apparent transition is observed at high temperatures, which appears to
satisfy scaling.  However, as pointed out by Mason and Kapitulnik, 
the apparent value of $\nu z
\approx 4/3$, rather than 7/3. Moreover, at low temperatures, the 
resistivity apparently
saturates to a finite (metallic) value, rather than diverging in
the putative insulating side of the transition or vanishing on the 
putative quantum Hall side.
The one striking difference between these results and the results for 
superconducting films is
that the apparent critical resistance at which the QHIT occurs 
appears to be, to good
approximation, universal (sample independent) and in agreement with 
the predictions of theory.

{\it A New Global Phase Diagram:}
Mason and Kapitulnik proposed figure 1b, where $\alpha$ is a measure
of dissipation.  While there are many speculative ideas concerning 
how this sort of phase
diagram could arise, at this stage this proposal must be viewed as
phenomenological.  In particular, our fundamental postulate that large $\alpha$
stabilizes a metallic phase is supported by recent theoretical work, 
especially that
in Ref. [24], but has not been clearly established. However, once this
postulate is accepted, it rationalizes in a simple way the 
observations on superconducting
films.  Moreover, on the basis of this idea, as outlined in the introduction,
a number of interesting further predictions can be made.

\acknowledgements{ We thank Assa Auerbach, Seb Doniach, Eduardo Fradkin,
Dung-Hai Lee, 
and Boris Spivak
for many useful discussions. Work at Stanford University supported by 
NSF grant DMR-9800663.
Work at UCLA was supported by NSF Grants DMR-98-14289 (SKA) and 
DMR-99-71138 (SC).
NM thanks Lucent CRFP for fellowship support.}


\begin{references}

\bibitem{sondhi}
For a review see, {\em e.g.},  S. L. Sondhi, S. M. Girvin, J. P. Carini,
and D. Shahar, Rev. Mod. Phys. 69, 315 (1997).

\bibitem{elihu}  
For a recent review of the two dimensional
MI-T in semiconductor MOSFETs and heterojunctions, see E.~Abrahams, 
S.~V.~Kravchenko, and
M.~P.~Sarachik, cond-mat/0006055.

\bibitem{chn}  
S.~Chakravarty, B.~I.~Halperin, D.~Nelson, \prl 60,
1057 (1988) and \prb 39, 2344 (1989).

\bibitem{subir}
S.~Sachdev, {\it Quantum Phase Transitions}, (Cambridge University Press,
2000) and references therein.

\bibitem{haviland}
D. B. Haviland {\it et al.} \prl 62, 2180 (1989); J. M. Valles, R.C.
Dynes, and J. P. Garno, \prl 69, 3567 (1992).  For a
review, see  e.g. Y.~Liu {\it et al}, \prb 47, 5931 (1993).

\bibitem{sudipandme}
S.~Chakravarty {\it et al}, \prl 56, 2303 (1986) and \prb (1987). 
M.~P.~A.~Fisher, \prl 36,
1917 (1987).

\bibitem{rimberg}
A. J. Rimberg, T. R. Ho, C. Kurdak, J. Clarke, K. L. Campman, and A. C.
Gossard, \prl 78, 2632 (1997).

\bibitem{takahide}
Y. Takahide, R. Yagi, A. Kanda, Y. Ootuka, and S. Kobayashi, \prl 85, 
1974 (2000).

\bibitem{tony}  
R. P. Feynman and F. L. Vernon JR., Ann. Phys. N.Y. 24,
118 (1963); K. Mohring and U. Smilansky, Nuc. Phys. A 338, 227 (1980);
A.O. Caldeira and A.J. Leggett, Ann. Phys. N.Y. 149, 374 (1984).

\bibitem{clastoquant}
S.~Chakravarty, \prl 49, 681 (1982); A.J. Bray and M. Moore, \prl 49,
1546 (1982); A. Schmid, \prl 51 1506 (1983); S. Chakravarty and A. J. Leggett,
\prl 52, 5 (1984). A. J. Leggett, S. Chakravarty, A. T. Dorsey, M. P. 
A. Fisher, A. Garg, and W. Zwerger, Rev. Mod. Phys. 59, 1 (1987).

\bibitem{mason1}
N. Mason and A. Kapitulnik, \prl 82, 5341 (1999).

\bibitem{trugman}
S.~Trugman, \prb 27, 7539 (1983).

\bibitem{chalker}
J. T. Chalker and P. D. Coddington, J. Phys. C 21, 2665 (1988).

\bibitem{lee1}
D. H. Lee, Z. Wang, and S. Kivelson, \prl 70, 4130 (1993).

\bibitem{klz}
S. Kivelson, D. H. Lee, and S.-C. Zhang, \prb 46, 2223 (1992).

\bibitem{zhk}
S.-C. Zhang, T.~H.~Hansson, and S.~A.~Kivelson,  \prl 62, 82(1990).

\bibitem{shahar}
D. Shahar, M. Hilke, C. C. Li, D. C. Tsui, S. L. Sondhi,
J. E. Cunningham, and M. Razeghi, Sol. St. Comm. 107, 19 (1998).

\bibitem{balaban}
N. Q. Balaban, U. Meirav, and I. Barjoseph, \prl 81, 4967 (1998).

\bibitem{zimanyi}
V.~J.~Emery and S.~A.~Kivelson, \prl 74, 3253 (1995);
K-H.~Wagenblast {\it et al}, \prl 79, 2730 (1997); K. Voelker, 
cond-mat/9911473.

\bibitem{randomflux}
Only long range correlated random flux model shows a metal-insulator
transition: D. N. Sheng and Z. Y. Weng, Europhys. Lett., 50, 776
(2000); H.~Nguyen and S.~Chakravarty, unpublished. The $\delta$-correlated
model shows only localized states. For theoretical work
on this model, see, {\em e.g.}, S-C.~Zhang and D. P. Arovas, \prl 72, 
1886 (1994); Y.~B.~Kim, A.~Furusaki, and D.~K.~K.~Lee,  \prb 52,  16646
(1995); A. Furusaki, \prl 82, 604 (1999).

\bibitem{sudipsideas}  
H.~Nguyen  and S.~Chakravarty, unpublished.

\bibitem{sak}
E. Shimshoni, A. Auerbach and A. Kapitulnik, \prl 80, 3352 (1998).

\bibitem{das}
D. Das and S. Doniach, \prb 60, 1261 (1999).

\bibitem{dalidovich}
D. Dalidovich and P. Phillips, cond-mat/0005119.

\bibitem{feigelman}
M. V. Feigelman and A. I. Larkin, Chemical Physics 235,  107 (1998);
cond-mat/9908075.

\bibitem{spivak1}
B. Spivak, A. Zyuzin, M. Hruska, cond-mat/0004058.

\bibitem{demler}
E. Demler, Y. Oreg and D. S. Fisher, preprint.

\bibitem{lerner}
I. V. Yurkevich, I. V. Lerner, cond-mat/0007317.

\bibitem{aalr}
E. Abrahams, P. W. Anderson, D. C. Licciardello, and T. V. Ramakrishnan,
\prl 42, 673 (1979)

\bibitem{aal}
B. L. Altshuler, A. G. Aronov, and P. A. Lee, \prl 44, 1288 (1980).

\bibitem{hebard}
A. F. Hebard and M. A. Paalanen, \prl 65, 927 (1990).

\bibitem{paalanen}
M.~A.~Paalanen {\it et al}, \prl 69, 1604 (1992).

\bibitem{yazdani}
A. Yazdani and A. Kapitulnik, \prl 74, 3037 (1995).

\bibitem{ephron}
D. Ephron, A. Yazdani, A. Kapitulnik, and M.R. Beasley, \prl 76, 1529 (1996).

\bibitem{valles}
J. A. Chervenak and J. M. Valles Jr., \prb 59, 11209 (1999).

\bibitem{zant}
H. S. J. van der Zant, W. J. Elion, L. J. Geerligs, and J. E. Mooij, 
\prb 54, 10081 (1996); C.D. Chen, P. Delsing, D. B. Haviland, Y. Harada, and T. 
Claeson, \prb 51, 16645 (1992).

\bibitem{mason2}
N. Mason and A. Kapitulnik, preprint 2000 (cond-mat/0006138).

\bibitem{liu}
Y. Liu, D. B. Haviland, L.I. Glazman, and A.M. Goldman, \prl 68, 2224 (1992).

\bibitem{dynes}
R. P. Barber, L. M. Merchant, A. Laporta, and R. C. Dynes, \prb 49,
3409 (1994); A. Frydman, L.M. Merchant, and R. C. Dynes, Physica 
Status Solidi B 218, 173 (2000).

\bibitem{goldman}
H. M. Jaeger, D. B. Haviland, B. G. Orr, and A. M. Goldman, \prb 40, 
182 (1989).

\bibitem{kravchenko}
S. V. Kravchenko, G. V. Kravchenko, J. E. Furneaux, V. M. Pudalov, and M.
Diorio,\prb 50, 8039 (1994);  S.~V.~Kravchenko and T.~M.~Klapwijk, 
\prl 84, 2909 (2000).

\bibitem{gold}
A. Gold, \prb 44, 8818 (1991).  A.~L.~Efros, \prl 68,2208 (1992).

\bibitem{jiang12b}
S. C. Dultz, H. W. Jiang, and W.  J. Schaff, \prb 58, R7532 (1998).

\bibitem{shaharnewun}
Y. Hanein, N. Nenadovic, D. Shahar, H. Shtrikman, I.Yoon, C.C. Li, 
and D.C. Tsui, Nature 400,
735 (1999).

\bibitem{simosfet}
A. A. Shashkin, G. V. Kravchenko, and V. T. Dolgopolov, JETP Lett. 58,
220 (1993)]; A. A. Shashkin, V. T. Dolgopolov, G. V. Ktavchenko, M.
Wendel, R. Schuster, J. P.  Kotthaus, R. J. Haug, K. von Klitzing, K. 
Ploog, H. Nickel, and W. Schlapp, \prl 73, 3141 (1994)

\bibitem{laughlin2}
R. B. Laughlin, \prb 23, 5632 (1981).

\bibitem{khmelnitskii}
D. E. Khmelnitskii, Helvetica Physica Acta 65, 164 (1992).

\bibitem{wong}
H.~W.~Jiang, C.~E.~Johnson, K.~L.~Wang, S.~T.~Hannahs, \prl 71, 1439 (1993);
I.~Glozman, C.~E.~Johnson, and H.~W.~Jiang, \prl 74, 594 (1995).

\bibitem{other}  
T.~Wang, K.~P.~Clark, G.~F.~Spencer, A.~M.~Mack, and W.~P.~Kirk, 
\prl 72, 709 (1994); R.~J.~F.~Hughes, J.~T.~Nicholls, J.~E.~F.~Frost, 
E.~H.~Linfield, M.~Pepper, C.~J.~B.~Ford,  D.~A.~Ritchie, G.~A.~C.~Jones, 
E.~Kogan, and M.~Kaveh, J. Phys. Cond. Matt. 6, 4763 (1994).

\bibitem{krav}  
S.~V.~Kravchenko, W.~Mason, J.~W.~Furneaux, and V.~M.~Pudalov,
\prl 75, 910 (1995).

\bibitem{song}  
S.-H.~Song, D.~Shahar, D.~C.~Tsui, Y.~H.~Xie, and D.~Monroe,
\prl 78, 2200 (1997);

\bibitem{jiang12}
H. W. Jiang, H. L. Stormer, D. C. Tsui, L. N. Pfeiffer, and K. W. 
West,  \prb 40, 12013 (1989).

\bibitem{half}
R. L. Willett, M. A. Paalanen, R. R. Ruel, K. W. West, L. N. Pfeiffer, and
D. J. Bishop, \prl 65, 112 (1990).

\bibitem{nu12}
see e.g. R. L. Willett, M. A. Paalanen, R. R. Ruel, K. W. West, L. N.
Pfeiffer, and D. J. Bishop, \prl 65, 112 (1990).

\bibitem{vgoldman}
L. P. Rokhinson and V. J. Goldman, \prb 56, R1672 (1997).

\bibitem{wong2}  
L.~W.~Wong, H.~W.~Jiang, and W.~J.~Schaff, \prb 54, 17323 (1996).

\bibitem{web}
P. Mohanty, E. M. Q. Jariwala, and R.A. Webb, \prl 78, 3366 (1997).

\bibitem{phasediverges}
For an earlier review see e.g D. J. Thouless in {\it Ill-Condensed
Matter}, Les Houches Summer School, R. Balian, R. Maynard, and G. Toulouse editors
(North-Holland, Amsterdam, 1979) p. 5; B. L. Altshuler, A. G. Aronov and 
D. E. Khemelnitskii, J. Phys. C 15, 7367 (1982); S. Chakravarty and A. Schmid, 
Phys. Rep. 140, 195 (1986); for a discussion of the recent controversy, 
see I. Aleiner, B. L. Altshuler and M. E. Gershenson, Waves in Random Media, 9, 201 (1999).

\bibitem{laughlin}
R. B. Laughlin and V. Kalmeyer, \prl 59, 2095 (1987); C.
Nayak, Phys. Rev. B {\bf 62}, 4880 (2000).

\bibitem{superconducting}
See, {\em e.g.} D. S. Fisher, M. P. A. Fisher, and D. A. Huse, \prb 43, 130
(1991).

\bibitem{braggglass}  
Although it is now fairly well established
(M.~Gingras and D .A.~Huse, \prb 53, 15193 (1996); S. Ryu, A.
Kapitulnik, and S. Doniach, \prl 77, 2300 (1996); C. Zeng,
P. L. Leath, and D. S. Fisher,\prl 82, 1935 (1999) ) that a true 
``Bragg glass'' phase (P.~Le Doussal and  T.~Giamarchi, \prb 52, 1242
(1995).) does not exist in two dimensions, under circumstances of
weak disorder and strong interactions
between vortices ({\it i.e.} at moderately large
$H$) that a state with exponentially rare dislocations in the vortex
lattice can occur, which would behave, to all practical purposes, 
like a Bragg glass phase.  In this case, a fairly sharp finite 
temperature crossover might occur which would strongly resemble 
a non-zero temperature phase transition to a superconducting state.

\bibitem{fisher1}
M. P. A. Fisher, \prl 65, 923 (1990).

\bibitem{hyman}
R.A. Hyman, M. Wallin, M. P. A. Fisher, S. M. Girvin, and A. P. Young,
\prb 51, 15304 (1995).

\bibitem{philmag}
S.~Chakravarty, S. Kivelson, C. Nayak, and K. Voelker, Phyl. Mag. B 
79, 859 (1999).

\bibitem{randomsinglet}
R. N. Bhatt and K. Yang, J. Appl. Phys. 83, 7231 (1998).

\bibitem{ferro}
C. Herring, Rev. Mod. Phys. 34, 631 (1962); D. J. Thouless, P. Phys.
Soc. Lond. 86, 893 (1965).

\bibitem{luhlier}
G. Misguich, C. Lhuillier, B. Bernu, and C. Waldtmann, \prb 60, 1064 (1999); C.
Lhuillier, B. Bernu, and G. Misguich, Int. J. Mod. Phys. B 13, 687 (1999);
G. Misguich, B. Bernu, C. Lhuillier, and C. Waldtmann, \prl 81, 1098 (1998).

\bibitem{rvb}
P.~W.~Anderson, Science 235, 1196 (1987);
S.~A.~Kivelson, D.~P.~Rokhsar, and J.~P.~Sethna, \prb 35, 8865 (1987);
D.~P.~Rokhsar and S.~A.~Kivelson, \prl 61, 2376 (1988);  R. Moessner and
S.~L.~Sondhi, cond-mat/0007378.

\bibitem{pairedinsulator}
In 3-D an insulating state has been observed in e.g. Y. Shapira and
G. Deutscher, Phys. Rev. B 27, 4463 (1983); for 2-D films see e.g. ref.
\cite{goldman}, and references therein.

\bibitem{orderoflimits}
In defining the appropriate dissipative
responses of the system, it is important to remember that the
thermodynamically meaningful conductivities  are
defined from the Kubo formula by taking limits in the following 
canonical order: first, the thermodynamic limit, then the limit $\vec k\rightarrow 0$,
then the limit $\omega\rightarrow 0$, and only after that,
the limit $T\rightarrow 0$.  Among other things, by taking the
zero temperature limit at the end, we presumably guarantee that
the mesocopic fluctuations of the conductivity can be ignored,
or in other words  that the disorder is self-averaging.  The resistivity
tensor is then the tensor inverse of the so defined conductivity tensor.

\bibitem{imry}
Y. Imry, \prl 71, 1868 (1993);O. Entin-Wohlman, A. G. Aronov, Y. 
Levinson, and Y. Imry, \prl 75, 4094 (1995).

\bibitem{quantizedinsul}
E. Shimshoni and A. Auerbach, \prb 55, 9817 (1997); Phil. Mag. B 77,
1107 (1998); D. Shahar, D.C. Tsui, M. Shayegan, J. E. Cunningham, 
E. Shimshoni, and S. L. Sondhi, Sol. St. Comm. 102, 817 (1997); 
M.~Hilke, D.~Shahar, S.~H.~Song, D.~C.~Tsui, Y.~H.~Xiet, D.~Monroe, 
Nature 395, 675 (1998).

\bibitem{fisherlee}
M.~P.~A.~Fisher and D.-H.~Lee, \prl 63, 903(1989).


\bibitem{jain}
J.~K.~Jain, \prl 64, 1297 (1990).

\bibitem{lopezfradkin}  
A.~Lopez and E.~Fradkin, \prb. 44, 5246
(1991); B.~I.~Halperin, P.~A.~Lee, and N.~Reed, \prb 47, 7312 (1993).


\bibitem{qhreview}  
A.~Karlhede, S.~A.~Kivelson, and S.~L.~Sondhi,in
{\it Correlated Electron Systems} ed. by V.~J.~Emery,  (World Scientific,
Singapore, 1993) pg. 242.

\bibitem{jtk}  
J.~K.~Jain, N.~Trivedi, and S.~A.~Kivelson,
\prl 64, 1297 (1990);  M.~Greiter and F.~Wilzcek, Nucl.
Phys. B 370, 577-600(1992).

\bibitem{caveatboson}  
Actually, even in the context of the dirty boson model,
there must be, in addition to the superconducting and insulating
phases, a variety of possible quantum Hall liquid phases. For example, it is
straightforward to show that, for an appropriate  model with short-range
repulsions between bosons, the bosonic $\nu=1/2$ Laughlin state is the exact
ground-state.  Thus, for example, it is in
principle possible to treat the $s_{xy}=1$ quantum Hall liquid as a
Bose condensate of $\theta=1$ Chern-Simons bosons, as a quantum Hall liquid of
$\theta=1+2n$ Chern-Simons bosons with $n\ne 0$, or as an insulating 
phase of the dual
Chern-Simons bosons with $\theta=-1$.

\bibitem{fisher2}
M. P. A. Fisher, G. Grinstein, and S. M. Girvin, \prl
64, 587 (1990); K. Kim and P. B. Weichman, \prb 43, 13583 (1991).

\bibitem{hsu}
S. Y. Hsu and J. M. Valles, \prb 49, 6416 (1994).

\bibitem{leewang}
D. H. Lee and Z. Q. Wang,  \prl 76, 4014 (1996);  Z.~Wang,
M.~P.~A.~Fisher, S.~M.~Girvin, J.~T.~ Chalker, \prb 61, 8326 (2000).

\bibitem{klk}
D. H. Lee, Y. Krotov, J. Gan, and S. A. Kivelson, \prb 57, 9349 (1997).

\bibitem{ruzin}  
A.~Dykhne and I.~Ruzin, \prb 50, 2369 (1994).

\bibitem{sondhi2} 
D.~Shahar, D.~C.~Tsui, M.~Shayegan, J.~E.~Cunningham,
E.~Shimshoni, and S.~L.~Sondhi, {\it Science} 274, 589 (1996);
E.~Shimshoni, S.~L.~Sondhi,  and D.~Shahar, \prb 55, 13730 (1997).

\bibitem{bhatt}  
Y.~Huo, R.~E.~Hetzel, and R.~N.~Bhatt, \prl 70, 481 (1993).

\bibitem{leeprivate}
D. H. Lee, private communication.

\bibitem{chlovskii}  
D.~B.~Chklovskii, B.~I.~Shklovskii, L.~I.~Glazman
\prb 46, 4026 (1992); D.~B.~Chklovskii and P.~A.~Lee, \prb 48, 18060 (1993).

\bibitem{as}
D. Stauffer and A. Aharony, {\it Introduction to Percolation Theory},
2nd ed.,  Taylor and Francis, London 1994.


\bibitem{lutken}  
C.~A.~Lutken and G.~G.~Ross, \prb 45, 11837 (1992) and \prb 48, 2500 (1993).

\bibitem{universality}
For a recent discussion of the issue of ``superuniversality'' of the quantum Hall transitions,
see L.~P.~Pryadko, \prb 56, 6810 (1999).  See also E.~Fradkin and
S.~A.~Kivelson, Nucl. Phys. B 474, 543 (1996); L.~P.~Pryadko and S.~C.Zhang,
\prb  54, 4953 (1996); J.~W.~Ye and S.~Sachdev \prl 80, 5409
(1998). See also  K.~Lutken and G.G. Ross \prb 48, 2500 (1993);
J.~K.~Jain \prl 64, 1297 (1990); N.~Trivedi, and S.~A.~Kivelson \prl 
64, 1297 (1990).

\bibitem{nonuniversal}
Critiques of the idea of superuniversality can be found in
X.G. Wen, \prb 46, 2655 (1992); X.G. Wen and Y.S. Wu, \prl 70, 1501
(1993); M. C. Cha, M. P. A. Fisher, S. M. Girvin, M. Wallin, and
A. P. Young, \prb 44, 6883 (1991).

\bibitem{danfisher}
D.~S. Fisher, \prl 69, 534 (1992); A.~P.~Young, Int. J. of Mod. Phys. 10, 1391 (1999); 
O.~Motrunich, {\it et al}, \prb 61, 1160 (2000); D.S. Fisher, Physica A 263, 222 (1999).

\bibitem{jiangqhit}
H.~W.~Jiang, C. E. Johnson, K. L. Wang, and S. T. Hannahs, \prl 71, 
1439 (1993); I. Glozman, C. E. Johnson, and H. W. Jiang, \prl 74, 594 (1995);
R. J. F. Hughes, J. T. Nicholls, J. E. F. Frost, E. H. Linfield,
M. Pepper, C. J. B. Ford, D. A. Ritchie, G. A.C. Jones, E. Kogan,
and M. Kaveh, J. Phys. C 6, 4763 (1994); T. Wang, K. P. Clark, G.
F. Spencer, A. M. Mack, and W. P. Kirk, \prl 72, 709 (1994).

\bibitem{trivedi}
L. W. Wong, H. W. Jiang, N. Trivedi, and E. Palm, \prb 51, 18033 (1995).

\bibitem{hilke}
M. Hilke, D. Shahar, S.H. Hong, D.C. Tsui, Y.H. Xie, and D. Monroe, 
\prb 56, 545 (1997)

\end{references}
\end{document}